\documentclass{birkjour}
 \newtheorem{thm}{Theorem}[section]
 \newtheorem{cor}[thm]{Corollary}
 
 \newtheorem{prop}[thm]{Proposition}
 \theoremstyle{definition}
 
 \theoremstyle{remark}

 \numberwithin{equation}{section}
 \usepackage{indentfirst}

\usepackage[all]{xy}

\title[Variational tricomplex and BRST theory]
{Variational tricomplex and BRST theory}
\author{Alexey A. Sharapov}
\address{Physics Faculty, Tomsk State University, Tomsk 634050, Russia}
\email{sharapov@phys.tsu.ru}

\begin{document}
\maketitle
\begin{abstract}
 By making use of the variational tricomplex,  a covariant procedure is proposed for deriving the classical BRST charge of the BFV formalism from a given BV master action.
\end{abstract}
\section{Introduction}

The BRST theory provides the most powerful approach to the  quantization of gauge systems \cite{HT}. It includes the Batalin-Vilkovisky (BV) formalism for Lagrangian gauge systems and its Hamiltonian counterpart known as the
Batalin-Fradkin-Vilkovisky (BFV) formalism. Usually, the two formalisms  are developed in parallel starting,
respectively, from the classical action or the first-class
constraints on the phase space of the system.
In either case one applies the homological
perturbation theory (hpt) to obtain the master action or the
classical BRST charge  at the output. A relationship between both the pictures of gauge dynamics is
established  through the Dirac-Bergmann (DB) algorithm, which allows
one to generate the complete set of first-class constraints by
the classical action. All these can be displayed diagrammatically as
follows:
$$
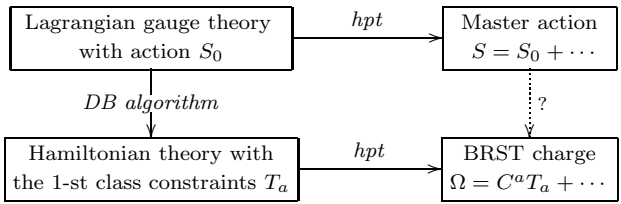
\xymatrix@R=10pt{*+[F]\txt{\Small \;Lagrangian gauge theory\;\;\\\Small with action $S_0$} \ar[rr]^-{\txt{\Small \textit{hpt}}}\ar[dd]|{\txt{\Small \textit{DB algorithm}}} &&*+[F]\txt{\Small Master action \\\Small \;\; $S=S_0+\cdots\;\;$}\ar@{.>}[dd]^{?}\\&&\\
*+[F]\txt{\Small Hamiltonian theory with\\\Small the $1$-st class constraints $T_a$} \ar[rr]^-{\txt{\Small \textit{hpt}}}&&*+[F]\txt{\Small BRST charge \\\Small $\Omega=C^aT_a+\cdots$} }
$$

Looking at this picture it is natural to ask about the dotted arrow making the diagram commute. The arrow symbolizes a hypothetical  map or construction connecting the BV and  BFV formalisms at the level of generating functionals. As we show below such a map really exists. By making use of the variational tricomplex \cite{Sh1}, we propose a direct construction of the classical BRST charge  from the BV master action. The construction is explicitly covariant (even though we pass to the Hamiltonian picture)  and generates the full spectrum of BFV ghosts  immediately from that of the BV theory. We also derive a covariant Poisson bracket on the extended phase space of the theory, with respect  to which the classical BRST charge obeys the master equation. The construction of the covariant Poisson bracket is similar to that presented in \cite{Dickey}, except that our Poisson bracket is defined off shell.

Finally, it should be noted that the first variational tricomplex for gauge systems was introduced in \cite{BH} as the Koszul-Tate resolution of the usual variational bicomplex for partial differential equations. Using this tricomplex, the authors of \cite{BH} were able to relate various Lie algebras associated with the global symmetries and conservation laws of a classical gauge system. Our tricomplex is similar in nature but involves the full BRST differential, and not its Koszul-Tate part.

\section{Variational tricomplex of a local gauge system}

In modern language the classical fields are just the sections of a locally trivial, fiber bundle $\pi : E\rightarrow M$ over an $n$-dimensional space-time manifold $M$. The typical fiber $F$ of $E$ is called the \textit{target space of fields}. In case the bundle is trivial, i.e., $E=M\times F$, the fields are merely  the smooth mappings from $M$ to $F$. For the sake of simplicity, we restrict ourselves to fields associated with vector bundles. In this case the space of fields $\Gamma (E)$ has the structure of a real vector space.

Bearing in mind gauge theories as well as field theories with fermions, we assume $\pi: E\rightarrow M$ to be a $\mathbb{Z}$-graded supervector bundle over the ordinary (non-graded) smooth manifold $M$. The Grassmann parity and the $\mathbb{Z}$-grading of a homogeneous  object $A$ will be denoted by  $\epsilon(A)$ and $\deg A$, respectively. It should be emphasized that in the presence of fermionic fields there is no natural correlation between the Grassmann parity and the $\mathbb{Z}\,$-grading. Since  throughout the paper we work exclusively in the category of $ \mathbb{Z}\,$-graded supermanifolds, we omit the boring prefixes ``super'' and ``graded'' whenever possible. For a quick introduction to the graded differential geometry and some of its applications we refer the reader to \cite{V1}, \cite{CatSch}, \cite{V2}.

In the local field theory, the dynamics of fields are governed by partial differential equations.
The best way to account for the local structure of fields is to introduce the  variational bicomplex $\Lambda^{\ast,\ast}(J^\infty E; d, \delta)$ on the infinite jet bundle $J^\infty E$ associated with the vector bundle $\pi: E\rightarrow M$. Here $d$ and $\delta$ denote the horizontal and vertical differentials in the bigraded space $\Lambda^{\ast,\ast}(J^\infty E)=\bigoplus \Lambda^{p,q}(J^\infty E)$ of differential forms on $J^\infty E$, where $p$ and $q$ refer  to the vertical and horizontal degrees, respectively. A brief account of the concept of a vatiational bicomplex can be found in \cite{Anderson}, \cite{Dickey}.

The free variational bicomplex represents thus a natural kinematical basis for defining local field theories.   In order to specify dynamics two more geometrical ingredients are needed. These are the classical BRST differential and the BRST-invariant (pre)symplectic structure on $J^\infty E$. Let us give the corresponding definitions.

\subsection{Presymplectic structure}\label{PrSt} By a \textit{presymplectic} $(2,m)$-form  on $J^\infty E$ we understand an
element $\omega\in \widetilde{\Lambda}^{2,m}(J^\infty E)$ satisfying
\footnote{By abuse of notation, we denote by $\omega$ an element of the quotient space $\widetilde{\Lambda}^{2,m}=\Lambda^{2,m}/d\Lambda^{2,m-1}$ and its representative   in $\Lambda^{2,m}$. The sign $\simeq$ means equality modulo $d\Lambda^{\ast,\ast}$.}
\begin{equation}\label{dom}
\delta \omega \simeq 0\,.
\end{equation}
The form $\omega$ is assumed to be homogeneous, so that we can speak of an odd or even  presymplectic structure of definite $\mathbb{Z}$-degree. The triviality of the relative ``$\delta$ modulo $d$'' cohomology in positive vertical degree (see \cite[Sec. 19.3.9]{Dickey}) implies that any presymplectic $(2,m)$-form is exact, namely, there exists a homogeneous  $(1,m)$-form $\theta$ such that $\omega\simeq\delta\theta$. The form $\theta$ is called the \textit{presymplectic potential} for $\omega$.  Clearly, the presymplectic potential is not unique. If $\theta_0$ is one of the presymplectic potentials for $\omega$, then setting   $\omega_0=\delta \theta_0$ we get
$$
\delta \omega_0=0\,,\qquad \omega_0\simeq \omega\,.
$$
In other words,  any presymplectic form has a  $\delta$-closed representative.

Denote by  $\ker\omega$ the space of all evolutionary vector fields $X$ on $J^\infty E$ that fulfill the relation\footnote{Recall that a vertical vector field $X$ is called \textit{evolutionary } if $i_X d+(-1)^{\epsilon({X})}di_{X}=0$, where $i_X$ is the operation of contraction of $X$ with differential forms.}
$$
i_X\omega \simeq 0\,.
$$
A presymplectic form $\omega$ is called non-degenerate if $\ker \omega=0$, in which case we refer to it as a \textit{symplectic form}.

An evolutionary vector field $X$ is called \textit{Hamiltonian} with respect to $\omega$ if it preserves the presymplectic form, that is,
\begin{equation}\label{Xom}
L_X\omega\simeq 0\,.
\end{equation}
Obviously, the Hamiltonian vector fields form a subalgebra in the Lie algebra of all evolutionary vector fields.  Eq. (\ref{Xom}) is equivalent to
$$
\delta i_X\omega\simeq 0\,.
$$
Again, because of the triviality of the relative $\delta$-cohomology, we can write
\begin{equation}\label{HVF}
i_X\omega \simeq \delta H
\end{equation}
for some $H\in \widetilde{\Lambda}^{0,m}(J^\infty E)$. We refer to $H$ as a \textit{Hamiltonian form} (or   \textit{Hamiltonian}) associated with $X$.  Sometimes, to indicate the relationship between the Hamiltonian vector fields and forms, we will write $X_H$ for $X$.  In general, the relationship is  far from being one-to-one.

The space ${\Lambda}_\omega^{0,m}(J^\infty E)$ of all Hamiltonian $m$-forms can be endowed with the structure of a Lie algebra. The corresponding Lie bracket is defined as follows: If $X_A$ and $X_B$ are two Hamiltonian vector fields associated with the Hamiltonian forms $A$ and $B$, then
\begin{equation}\label{PB}
\{A,B\}=(-1)^{\epsilon(X_A)}i_{X_A}i_{X_B}\omega\,.
\end{equation}
The next proposition shows that the bracket is well defined and possesses  all the required properties.

\begin{prop}[\cite{Sh1}]\label{2.1}
The bracket (\ref{PB}) is bilinear over reals, maps the Hamiltonian forms to Hamiltonian ones, enjoys the symmetry property
\begin{equation}\label{sym}
\{A,B\}\simeq -(-1)^{(\epsilon(A)+\epsilon(\omega))(\epsilon(B)+\epsilon(\omega))}\{B,A\}\,,
\end{equation}
and obeys  the Jacobi identity
\begin{equation}\label{jac}
\{C,\{A,B\}\}\simeq \{\{C,A\},B\}+(-1)^{(\epsilon(C)+\epsilon(\omega))(\epsilon(A)+\epsilon(\omega))}\{A,\{C,B\}\}\,.
\end{equation}
\end{prop}

\subsection{Classical BRST differential}\label{brst}
An odd evolutionary  vector field $Q$ on $J^\infty E$ is called \textit{homological} if
\begin{equation}\label{QQ}
[Q,Q]=2Q^2=0\,, \qquad \deg\,Q=1\,.
\end{equation}
The Lie derivative along the homological vector field $Q$ will be denoted by $\delta_Q$. It follows from the definition that $\delta_Q^2=0$. Hence, $\delta_Q$ is a differential of the algebra $\Lambda^{\ast,\ast}(J^\infty E)$ increasing the $\mathbb{Z}$-degree by 1. Moreover, the operator $\delta_Q$ anticommutes with the coboundary operators $d$ and $\delta$:
$$
\delta_Q d+d\delta_Q=0\,,\qquad \delta_Q \delta+\delta\delta_Q=0\,.
$$
This allows us to speak of the tricomplex $\Lambda^{\ast,\ast,\ast}(J^\infty E; d, \delta, \delta_Q)$, where
$$
\delta_Q: \Lambda^{p,q,r}(J^\infty E)\rightarrow \Lambda^{p,q,r+1}(J^\infty E)\,.
$$

 In the physical literature the homological vector field $Q$ is known as the \textit{classical BRST differential}  and the $\mathbb{Z}$-grading is called the \textit{ghost number}. These are the two main ingredients of all modern approaches to the covariant quantization of gauge theories.  In the BV formalism, for example, the BRST differential carries all the information about equations of motions, their  gauge symmetries and identities, and the space of physical observables is naturally identified with the group  $H^{0,{n},0}(J^\infty E; \delta_Q/d) $ of ``$\delta_Q$ modulo $d$'' cohomology in ghost number zero. For general non-Lagrangian gauge theories the classical BRST differential was systematically defined in \cite{LS0}, \cite{KazLS}.

The equations of motion of a gauge theory can be recovered by considering the zero locus of the homological vector field $Q$. In terms of adapted coordinates $(x^i, \phi^a_I)$ on $J^\infty E$ the vector  field $Q$, being evolutionary, assumes the form\footnote{We use the multi-index notation according to which the multi-index $I=i_1i_2\cdots i_k$ represents the set of symmetric covariant indices and $\partial_I=\partial_{i_1}\cdots\partial_{i_k}$. The \textit{order} of the multi-index is given by $|I|=k$.}
$$
Q=\partial_I Q^a\frac{\partial}{\partial \phi_I^a}\,.
$$
Then there exists an integer $l$ such that the equations
$$
\partial_I Q^a=0\,,\qquad |I|=k\,,
$$
define a submanifold $\Sigma^k\subset J^{l+k}E$. The standard regularity condition implies that $\Sigma^{k+1}$ fibers over $\Sigma^k$ for each $k$. This gives the infinite sequence of projections

$$
\xymatrix{\cdots\ar[r]& \Sigma^{l+3}\ar[r]&\Sigma^{l+2}\ar[r]&\Sigma^{l+1}\ar[r]&\Sigma^l}\rightarrow M\,,
$$
which enables us to define the zero locus of $Q$ as the inverse limit
 $$
 \Sigma^\infty =\lim_{\longleftarrow}\Sigma^k\,.
 $$
In physics, the submanifold $\Sigma^\infty\subset J^\infty E$ is usually referred to as  the \textit{shell}. The terminology is justified by the fact that the classical field equations as well as their differential consequences can be written as
$$
(j^{\infty}\phi)^\ast (\partial_I Q^a)=0\,.
$$
In other words, the field $\phi\in \Gamma(E)$ satisfies the classical equations of motion  iff $j^\infty \phi \in \Sigma^\infty$.

\subsection{$Q$-invariant presymplectic structure and its descendants}
By a \textit{gauge system} on $J^\infty E$ we will mean a pair $(Q, \omega)$ consisting of a homological vector field $Q$ and a $Q$-invariant presymplectic $(2,m)$-form $\omega$. In other words, the vector field $Q$ is supposed to be Hamiltonian with respect to $\omega$, so that $\delta_Q\omega\simeq 0$. The last relation implies the existence  of  forms $\omega_1$, $H$, and $\theta_1$ such that
\begin{equation}\label{des}
\delta_Q \omega=d\omega_1\,, \qquad   i_Q\omega =\delta H +d\theta_1\,.
\end{equation}
 As was mentioned in Sec.\ref{PrSt}, we can always assume that $\omega =\delta\theta$ for some presymplectic potential $\theta$, so that  $\delta\omega=0$. Then applying $\delta$ to the second equality in (\ref{des}) and using the first one, we find $d(\omega_1-\delta\theta_1)=0$. On account of the exactness of the variational bicomplex, the last relation is equivalent to
$$
\omega_1\simeq \delta\theta_1\,.
$$
Thus, $\omega_1$ is a presymplectic $(2,m-1)$-form on $J^\infty E$ coming from the presymplectic potential $\theta_1$. Furthermore, the form $\omega_1$ is $Q$-invariant as one can easily see by applying $\delta_Q$ to the first equality in (\ref{des}) and using once again the fact of exactness of the variational bicomplex. Let $H_1$ denote the Hamiltonian for $Q$ with respect to $\omega_1$, i.e.,
$$
i_Q\omega_1\simeq \delta H_1\,, \qquad H_1\in \widetilde{\Lambda}^{0,m-1}(J^\infty E)\,.
$$
Given the pair $(Q,\omega)$, we call $\omega_1$ the \textit{descendent presymplectic structure} on $J^\infty E$ and refer to $(Q,\omega_1)$ as the \textit{descendent gauge system}.

The next proposition provides an alternative definition for the descendent Hamiltonian of the homological vector field.
\begin{prop}[\cite{Sh1}]\label{p2}
Let $\omega$ be a $\delta$-closed representative of a presymplectic $(2,m)$-form on $J^\infty E$ and $\mathrm{deg} H_1\neq 0$, then
\begin{equation}\label{HH}
dH_1=-\frac12\{H,H\}\,.
\end{equation}
\end{prop}

\begin{cor}\label{c1}
$H$ is a Maurer-Cartan element of the Lie algebra $\Lambda^{0,m}_\omega(J^\infty E)$, that is, $$\{H,H\}\simeq 0\,.$$
\end{cor}

\begin{cor}\label{CL}
The Hamiltonian form $H_1$ is $d$-closed on-shell. In particular, for $m=n$ it defines a conservation law.
\end{cor}

\begin{prop}[\cite{Sh1}]\label{p5}
Suppose that the $Q$-invariant presymplectic form $\omega$ of top horizontal degree has the structure
\begin{equation}\label{ff}
\omega=P_{ab}\wedge\delta\phi^a\wedge\delta\phi^b\,,\qquad P_{ab}\in \Lambda^{0,n}(J^\infty E)\,,
\end{equation}
and $H$ is the Hamiltonian of $Q$ with respect to $\omega$. Then the presymplectic potential for the descendent presymplectic (2,n-1)-form  $\omega_1\simeq \delta\theta_1$ is defined  by the equation
\begin{equation}\label{ht1}
\delta H=\delta\phi^a\wedge \frac{\delta H}{\delta\phi^a}-d\theta_1 \,.
\end{equation}
\end{prop}

The above construction of the descendent gauge system $(Q,\omega_1)$ can be iterated producing a sequence of gauge systems $(Q, \omega_k)$, where the $k$-th presymlectic form $\omega_k\in {\Lambda}^{2,m-k}(J^\infty E)$ is the descendant of $\omega_{k-1}$. The minimal $k$ for which $\omega_k \simeq 0$ gives a numerical  invariant of the original gauge system $(Q,\omega)$.

\section{BFV from BV}\label{BV-BFV}

In this section, we apply the construction of the variational tricomplex for establishing  a direct correspondence  between the BV formalism of Lagrangian gauge systems and its Hamiltonian counterpart known as the BFV formalism. We start from a very brief account of both the formalisms in a form suitable for our purposes. For a systematic exposition of the subject we refer the reader to \cite{HT}.

\subsection{BV formalism} The starting point of the BV formalism is an infinite-dimensional manifold $\mathcal{M}_0$ of gauge fields that live on an $n$-dimensional space-time $M$. Depending on a particular structure of gauge symmetry the manifold $\mathcal{M}_0$ is extended to an $\mathbb{N}$-graded manifold $\mathcal{M}$ containing $\mathcal{M}_0$ as its body. The new fields of positive $\mathbb{N}$-degree are called the \textit{ghosts} and the $\mathbb{N}$-grading is referred to as the \textit{ghost number}. Let us collectively  denote all the original fields and ghosts by $\Phi^A$ and refer to them as fields. At the next step the space of fields $\mathcal{M}$ is further extended  by introducing the odd cotangent bundle $\Pi T^\ast[-1]\mathcal{M}$. The fiber coordinates, called \textit{antifields}, are denoted by $\Phi_A^\ast$. These are assigned with the following ghost numbers and Grassmann parities:
$$
\mathrm{gh} (\Phi^\ast_A)=-\mathrm{gh} (\Phi^A)-1\,,\qquad \epsilon (\Phi^\ast_A)=\epsilon (\Phi^A)+1 \quad (\mbox{mod}\, 2)\,.
$$
Thus, the total space of the odd cotangent bundle $\Pi T^\ast[-1]\mathcal{M}$ becomes a $\mathbb{Z}$-graded supermanifold.  The canonical Poisson structure on $\Pi T^\ast[-1]\mathcal{M}$ is determined  by  the following odd Poisson bracket in the space of functionals of $\Phi$ and $\Phi^\ast$:
\begin{equation}\label{abr}
(A,B)=\int_M \left(\frac{\delta_r A}{\delta \Phi^A}\frac{\delta_l B}{\delta \Phi^\ast_A}-\frac{\delta_r A}{\delta \Phi^\ast_A}\frac{\delta_l B}{\delta \Phi^A}\right)d^nx\,.
\end{equation}
 Here $d^nx$ is a volume form on $M$ and the subscripts $l$ and $r$ refer to the standard left and right functional derivatives.
In the physical literature the above bracket  is usually called the \textit{antibracket} or the \textit{BV bracket}.

The functionals of the form
$$
A=\int_M (j^\infty \phi)^\ast(a)\,,
$$
where $\phi=(\Phi, \Phi^\ast)$ and $a\in \widetilde{\Lambda}^{0,n}(J^\infty E)$, are called \textit{local}. Under suitable boundary conditions for $\phi$'s the map $a \mapsto A$ defines an isomorphism of vector spaces, which gives rise to a pulled-back Poisson bracket on $\widetilde{\Lambda}^{0,n}(J^\infty E)$. This last bracket is determined  by the symplectic structure
\begin{equation}\label{ops}
\omega= \delta \Phi_A^\ast\wedge \delta\Phi^A\wedge d^nx
\end{equation}
according to (\ref{PB}).
By definition, $\mathrm{gh} (\omega)= - 1$ and $\epsilon (\omega)=1$.

The central goal of the BV formalism is the construction of a \textit{master action} $S$ on the space of fields and antifields. This is defined as a proper solution to the \textit{classical master equation}
\begin{equation}\label{BV_MEq}
(S,S)=0\,.
\end{equation}
The local functional $S$ is required to be of ghost number zero and start with the action $S_0$ of the original fields to which one couples vertices involving antifields. All these vertices can be found systematically from the master equation (\ref{BV_MEq}) by means of the so-called \textit{homological perturbation theory} \cite{HT}.

The classical BRST differential on the space of fields and antifields is canonically generated by the master action through the antibracket:
\begin{equation}\label{CLBRSTD}
Q=(S\,,\,\cdot\,)\,.
\end{equation}
Because of the master equation for $S$ and the Jacobi identity for the antibracket (\ref{abr}),  the operator $Q$ squares to zero in the space of smooth functionals.  The physical quantities are then identified with the cohomology classes of $Q$ in ghost number zero. When restricted to the subspace of local functionals the classical BRST differential (\ref{CLBRSTD}) induces a homological vector field on the total space of the jet bundle $J^\infty E$.

\subsection{BFV formalism} The Hamiltonian formulation of the same gauge dynamics implies a prior splitting $M=N\times \mathbb{R}$ of the original space-time into space and time; the factor $N$ can be viewed as the physical space at a given instant of time. The initial values of the original fields are then considered to form an infinite-dimensional manifold $\mathcal{N}_0$. To allow for possible constraints on the initial data of fields the manifold  $\mathcal{N}_0$ is extended to an $\mathbb{N}$-graded supermanifold $\mathcal{N}$ by adding new fields, called ghosts, of positive $\mathbb{N}$-degree. Then the space of original fields and ghosts is doubled by introducing  the cotangent bundle $T^\ast \mathcal{N}$ endowed with the canonical symplectic structure.     If we denote the local coordinates on $\mathcal{N}$ by $\Phi^a$ and the linear coordinates in the cotangent spaces by $\bar{\Phi}_a$, then the canonical Poisson bracket in the space of functionals of $\Phi^a$ and $\bar{\Phi}_a$ reads
\begin{equation}\label{epb}
\{A,B\}=\int_N \left(\frac{\delta_r A}{\delta \Phi^a}\frac{\delta_l B}{\delta \bar{\Phi}_a}-(-1)^{\epsilon(\Phi_a)}\frac{\delta_r A}{\delta \bar{\Phi}_a}\frac{\delta_l B}{\delta \Phi^a}\right)d^{n-1}x\,.
\end{equation}
Here $d^{n-1}x$ stands for a volume form on $N$. By the definition of the cotangent bundle of a graded manifold
$$
{\mathrm{gh}} ( \bar{\Phi}_a) =-\mathrm{gh}({\Phi}^a)\,,\qquad \epsilon(\bar{\Phi}_a) =\epsilon({\Phi}^a)\,,
$$
Again, the space of local functionals, i.e., functionals of the form
 $$
 B=\int_N j^\infty(\phi)^\ast (b)\,, \qquad \phi=(\Phi,\bar\Phi)\,,\qquad b\in \tilde{\Lambda}^{0,n-1}(J^\infty E)\,,
 $$
 appears to be closed w.r.t. the even Poisson bracket (\ref{epb}) and  the map $b\mapsto B$ induces an even Poisson bracket on $\widetilde{\Lambda}^{0,n-1}(J^\infty E)$. The latter is determined by the even symplectic form
$$
\omega_1=\delta \bar{\Phi}_a\wedge \delta \Phi^a \wedge d^{n-1}x
$$
of ghost number zero.

 The gauge structure of the original dynamics is encoded by the \textit{classical $BRST$ charge} $\Omega$. This is given by an odd, local functional of ghost number $1$ satisfying the classical master equation
$$
\{\Omega,\Omega\}=0\,.
$$
The classical BRST differential in the
extended space of fields and momenta is given now by the Hamiltonian action of the BRST charge:
\begin{equation}\label{BDIF}
Q=\{\Omega\,,\,\cdot\,\}\,.
\end{equation}
It is clear that $Q^2=0$. The group of $Q$-cohomology in ghost number zero is then naturally identified with the space of physical observables. Upon restriction to the space of local functionals the variational vector field (\ref{BDIF}) induces a homological vector field on the total space of the infinite jet bundle.

\subsection{From BV to BFV} It must be clear from the discussion above that any gauge system  in the BFV formalism may be viewed as the descendant of the same system in the BV formalism. More precisely, we can define the even presymplectic structure $\omega_1$ on the phase space of a gauge theory  as the  descendant  of the odd symplectic structure (\ref{ops}):
$$
d\omega_1 =\delta_Q(\delta \Phi^\ast_A\wedge \delta \Phi^A\wedge d^nx)=\delta \left( \delta\Phi^A\wedge \frac{\delta S}{\delta \Phi^A}+\delta\Phi^\ast_A\wedge \frac{\delta S}{\delta \Phi^\ast_A}\right)\,.
$$
The corresponding classical BRST charge is given by
$$
\Omega_N=\int_N (j^{\infty}\phi)^\ast(J)\,,
$$
where $N\subset M$ is a space-like, Cauchy hypersurface and $J\in \Lambda^{0,n-1}_{\omega_1}(J^\infty E)$  is the Hamiltonian of the classical BRST differential $Q=(S,\,\cdot\,)$ w.r.t. the descendent presymplectic form $\omega_1$, i.e.,
\begin{equation}\label{J}
\delta J\simeq i_Q\omega_1\,.
\end{equation}
It is clear that $\mathrm{gh}(\Omega)=1$. In virtue of Corollary \ref{c1}, the functional $\Omega$ obeys the classical master equation $\{\Omega,\Omega\}=0$ with respect to the even Poisson bracket associated with $\omega_1$. According to Corollary \ref{CL} the form $J$ represents a conserved current, the BRST current. Formally, this means that the ``value'' of the odd charge $\Omega_N$ does not depend on the choice of $N$ provided that $j^\infty\phi\in \Sigma^\infty$.

Since the canonical symplectic structure (\ref{ops}) on the space of fields and antifields is $\delta$-exact, we can give an equivalent definition for $J$ in terms of the antibracket (\ref{abr}).  For this end, consider the dynamics of fields in a domain $D\subset M$ bounded by  two Cauchy hypersurfaces $N_1$ and $N_2$. The fields and antifields are assumed to vanish on space infinity together with their derivatives. By Proposition \ref{p2},
$$
-\frac12(S,S)=\int_D (j^\infty \phi)^\ast (d J)=\int_D d [(j^\infty \phi)^\ast (J)]=\Omega_{N_2}-\Omega_{N_1}\,.
$$

Let us illustrate the general construction by a particular example of gauge theory.

\subsection{Maxwell's electrodynamics} In the BV formalism, the free electromagnetic field in $4$-dimensional Minkowski space is described by the master action
\begin{equation}\label{MED}
  S=\int L   \,,\qquad L=-\Big(\frac14 F_{\mu\nu}F^{\mu\nu}+C\partial^\mu A_\mu^\ast\Big)d^4x\,.
\end{equation}
Here
$$
F_{\mu\nu}=\partial_\mu A_\nu-\partial_\nu A_\mu
$$
is the strength tensor of the electromagnetic field, $A^\ast_\mu$ is the antifield to the electromagnetic potential $A_\mu$, and $C$ is the ghost field associated with the standard gauge transformation
$
\delta_\varepsilon A_\mu=\partial_\mu \varepsilon
$.

Since the gauge symmetry is abelian, the master action (\ref{MED}) does not involve the ghost antifield $C^\ast$. The odd symplectic structure (\ref{ops}) on the space of fields and antifields assumes the form
$$
\omega=(\delta A^\ast_\mu\wedge \delta A^\mu+\delta C^\ast\wedge \delta C)\wedge d^4 x\,,\qquad d^4x=dx^0\wedge dx^1\wedge dx^2\wedge dx^3\,,
$$
and the action of the classical BRST differential is given by
\begin{equation}\label{brst-t}
\delta_Q A_\mu =\partial_\mu C\,,\qquad \delta_Q A^\ast_\mu =\partial^\nu F_{\nu\mu}\,,\qquad \delta_Q C=0\,, \qquad \delta_Q C^\ast=\partial^\mu A^\ast_\mu\,.
\end{equation}
The variation of the Lagrangian  density reads
$$
\delta L=(\partial^\mu F_{\mu\nu}\delta A^\nu +  \partial^\mu A^\ast_\mu\delta C +\partial^\mu C\delta A^\ast_\mu  -\partial^\mu \theta_\mu)\wedge d^4x\,,\qquad \theta_\mu =
F_{\mu\nu}\delta A^\nu +C \delta A_\mu^\ast\,.
$$
One can easily check that $i_Q \omega \simeq \delta L$.  By Proposition \ref{p5} the form
$$
\theta_1=-\theta_\mu\wedge d^3x^\mu\,,\qquad d^3x^\mu=\eta^{\mu\nu}i_{\frac{\partial}{\partial x^\nu}}d^4x\,,$$
defines the potential for the descendent presymplectic form
\begin{equation}\label{WC}
\omega_1=\delta\theta_1= -(\delta F_{\nu\mu}\wedge \delta A^\mu+\delta C\wedge \delta A^\ast_\nu)\wedge d^3x^\nu\,.
\end{equation}
(Of course, one  could arrive at this expression by considering the BRST variation $\delta_Q\omega = d\omega_1$ of the original symplectic structure.)

Applying the BRST differential to the form $\omega_1$ yields one more descendent presymplectic form
$$
\omega_2= \delta C\wedge \delta F_{\mu\nu}\wedge d^2x^{\mu\nu}\,, \qquad d^2x^{\mu\nu}=\eta^{\mu\alpha}i_{\frac{\partial}{\partial x^\alpha}}d^3x^\nu\,.
$$
This last form, being ``absolutely'' invariant under the BRST transformations (\ref{brst-t}), leaves no further descendants.

The $3$-form of the conserved BRST current $J$ associated to the BRST symmetry transformations (\ref{brst-t}) is determined by Eq. (\ref{J}). We find
$$
J=J_\nu d^3x^\nu\simeq-C\partial^\mu F_{\mu\nu} d^3x^\nu\,.
$$
 Once we identify $x^0$ with time in the Hamiltonian formalism, the antifield $A^\ast_0$ plays the role of ghost momentum canonically conjugate to $C$ with respect to the presymplectic structure (\ref{WC}). The on-shell conservation of the corresponding BRST charge $\Omega=\int_{\mathbb{R}^3} J_0d^3x$ expresses  nothing but the Gauss law $\partial^i F_{i0}=0$.

\vspace{0.2 cm}

\noindent \textbf{Acknowledgements.} The work was partially supported by the
RFBR grant 13-02-00551.

\end{document}